\documentclass{article}
\usepackage{fleqn}
\usepackage{epsfig}
\usepackage{amsmath,amssymb}
\newcommand{\Const}{\mathop{\rm Const}\nolimits}

\begin{document}

\begin{center}
{\bf\Large Collisionless self-gravitating statistical systems of scalar interacting particles}\\[12pt]
Yu.G. Ignat'ev\\
Kazan Federal University,\\ Kremlyovskaya str., 35,
Kazan 420008, Russia
\end{center}

\begin{abstract}
This paper is devoted to consideration of the theory of collisionless statistical systems with interparticle scalar interaction. The mathematical model of such systems is constructed and the exact solution of Vlasov equation for isotropic homogenous model of the Universe is found. Asymptotic solutions of self-consistent Vlasov - Einstein model for conformally invariant scalar interactions are found.
\end{abstract}

{\bf keywords}: Relativistic Kinetics, Scalar Fields, Scalar Interaction of Particles, Collisionless System.\\
{\bf PACS}: 04.20.Cv, 98.80.Cq, 96.50.S  52.27.Ny

\section{Introduction}
\label{intro}
In series of Author's works \cite{Ignatev3}, \cite{YuNewScalar3}, \cite{Yu_stfi14} etc. the mathematical model of self-gravitating statistical system of scalar charged particles is formulated. In particular, in \cite{GC4_15} this model is extended on sector of negative effective masses of scalar charged particles. However, in these papers as well as in other works of the Author with his pupils it was usually considered a locally equilibrium system. As is well-known, at inflationary stage thermodynamic equilibrium is violated \cite{Yu_LTE} hence it is reasonable to consider collisionless situation realized at this stage which is described by corresponding Vlasov equations.
This paper is devoted to this problem.

\section{Self-consistent Mathematical Model of Collisionless Plasma of Scalar Charged Particles}
\subsection{Collisionless Kinetic and Transport Equations}

Distribution functions $f_a(x,P)$ of particles of ``$a$'' sort having scalar charge $q_a$ are defined by invariant collisionless kinetic equations \cite{Ignatev3}\footnote{i.e. Vlasov equations.}:
\begin{equation} \label{GrindEQ__52_}
[H_a,f_a]=0,
\end{equation}
where $[H_a,f_a]$ is a Poisson bracket which can be written using covariant Cartan operator of differentiation $\widetilde{\nabla}$
in form\footnote{Details see in \cite{GC4_15}}:
\begin{equation}\label{H_Cart}
[H,\Psi ]\equiv \frac{\partial H}{\partial P_{i}} \widetilde{\nabla}_i\Psi-\frac{\partial \Psi}{\partial P_{i}} \widetilde{\nabla}_i H,
\end{equation}
Hamilton function of a particle in scalar field and normalization ratio for generalized momentum have the next form:
\begin{equation}\label{H,m}
H(x,P)=\frac{1}{2} \left[m_*^{-1}(x)(P,P)-m_*\right]=0,
\end{equation}
\begin{equation}\label{P_norm}
(P,P)=m^2_*\equiv (q_qa\Phi)^2.
\end{equation}

Distribution functions $f_a(x,P)$ of particles of ``$a$''-sort having scalar charge $q_a$ are defined through invariant collisionless kinetic equations \cite{Ignatev3}:
\begin{equation} \label{kin_eq}
[H_a,f_a]=0,
\end{equation}
where Hamilton function and normalization ratio for generalized momentum have the next form:
\begin{equation}\label{H,m}
H(x,P)=\frac{1}{2} \left[m_*^{-1}(x)(P,P)-m_*\right]=0,
\end{equation}
\begin{equation}\label{P_norm}
(P,P)=m^2_*\equiv (q_qa\Phi)^2.
\end{equation}
\begin{equation}\label{H_Cart}
[H,\Psi ]\equiv \frac{\partial H}{\partial P_{i}} \widetilde{\nabla}_i\Psi-\frac{\partial \Psi}{\partial P_{i}} \widetilde{\nabla}_i H,
\end{equation}

Let us notice identity laws, valid for Hamilton function \eqref{H,m} which will be useful in future:
\begin{equation}\label{nabla_H}
\widetilde{\nabla}_iH=-\nabla_i m_*,
\end{equation}
\begin{equation}\label{HPsi}
[H,\Psi]=\frac{1}{m_*}P^i\widetilde{\nabla}_i\Psi+\partial_i m_*\frac{\partial \Psi}{\partial P_i},
\end{equation}
where $\Psi(x,P)$ is an arbitrary function.

Further, let us define macroscopic moments using distribution functions:
\begin{equation} \label{cons_n}
n^{i}_a =\frac{2S+1}{(2\pi )^{3} } \; g_{a}
\int\limits_{P_0} f_a(x,P)P^{i} dP_0.
\end{equation}
- particle number density vectors which are identically conserved in consequence of kinetic equations
(\ref{kin_eq});
plasma energy-momentum tensor
\begin{equation}\label{Tpl}
T^{ik}_p=\sum\limits_{a} \frac{2S+1}{(2\pi )^{3}}
\int\limits_{P_0} f_a(x,P)P^iP^kdP_0
\end{equation}
and plasma scalar charge density
\begin{equation}\label{s}
\sigma=\Phi\sum\limits_a \frac{2S+1}{(2\pi )^{3} } q^2_a
\int\limits_{P_0} f_a(x,P)dP_0.
\end{equation}
These macroscopic moments are related to each other by transport equations (energy-momentum conservation law).

\begin{equation} \label{transport_Tik}
\nabla _{k} T_{p}^{ik} -\sigma\nabla ^{i} \Phi =0.
\end{equation}
\subsection{Scalar Field Equation}
First, let us consider Lagrangian function for a classical massive real scalar field $\Phi$. In such a case the Lagrangian of real scalar field can be chosen in the next form:
\begin{equation} \label{Ls} L_{s} =\frac{\epsilon_1}{8\pi } \left(g^{ik} \Phi _{,i} \Phi _{,k} -\epsilon_2 m_{s}^{2} \Phi ^{2} \right), \end{equation}
where $m_{s} $ is a mass of scalar field's quanta and it is $\epsilon_2=1$ for a classical scalar field,
$\epsilon_2=-1$ for a phantom scalar field; $\epsilon_1=1$ for a field with repulsion of like-charged particles, %
$\epsilon_1=-1$ for a field with attraction of like-charged particles $\epsilon_1=-1$.
Then energy-momentum tensor of a scalar field is:
\begin{equation} \label{Tiks} T_{s}^{ik} =\frac{\epsilon_1}{8\pi } \left(2\Phi ^{,i} \Phi ^{,k} -
g^{ik} \Phi ^{,j} \Phi _{,j} +\epsilon_2 g^{ik} m_{s}^{2} \Phi ^{2} \right). \end{equation}
Let us write the equation of massive non-conformal scalar field with a source \cite{Ignatev3}:
\begin{equation} \label{EqPhi}{\rm \square }\Phi +m_{s}^{2} \Phi =-4\pi \epsilon_1 \sigma,
\end{equation}
where
\[{\rm \square }\Phi \equiv g^{ik} \nabla _{i} \nabla _{k} \Phi =\frac{1}{\sqrt{-g} } \frac{\partial }{\partial x^{i} } \sqrt{-g} g^{ik} \frac{\partial }{\partial x^{k} } \Phi \]

Let us consider now the Lagrangian function for a classical massive real conformal scalar field  $\Phi $ (see e.g., [4,5];
for a massive scalar field the conformal invariance is understood as asymptotic property at ($m_{s} \to 0$)):
\begin{equation} \label{LsR} L_{s} =\frac{\epsilon_1}{8\pi } \left(g^{ik} \Phi _{,i} \Phi _{,k}  -\epsilon_2 m_{s}^{2} \Phi ^{2} +
\frac{R}{6}\Phi ^{2} \right). \end{equation}
The Lagrangian function differs from the standard one (see e.g. \cite{Melnikov})
in presence of factor $1/8\pi$ as well as introduced unit indicators $\epsilon_\alpha$. In addition, Ricci tensor in the article is obtained by means of convolution of first and third indices of Riemann tensor $R_{jl}=g^{ik}R_{ijkl}$. Components of energy-momentum tensor of a scalar field relative to Lagrangian function \eqref{LsR} are equal to \cite{Melnikov,Yubook1}:

\begin{eqnarray}
\label{TiksR}
\hskip -12pt T_{s}^{ik}=\frac{\epsilon_1}{8\pi } \biggl[\frac{4}{3}\Phi^{,i}\Phi ^{,k}\!\!-\!\!\frac{1}{3}g^{ik}\Phi_{,j}\Phi^{,j}+\epsilon_2 m^2_sg^{ik}\Phi^2+\nonumber\\
\hskip -12pt \frac{1}{3} \bigl(R^{ik} -\frac{1}{2} Rg^{ik} \bigr)\Phi ^{2}-\frac{2}{3}\Phi\Phi^{,ik}+\frac{2}{3}g^{ik}\Phi\Box\Phi\biggr].
\end{eqnarray}
assuming $\Phi \not\equiv {\rm Const}$, we get equation of massive scalar field with a source (see \cite{Yubook1}):
\begin{equation} \label{EqPhiR} {\rm \square }\Phi +\epsilon_2 m_{s}^{2} \Phi -\frac{R}{6} \Phi =
-4\pi \epsilon_1 \sigma.
\end{equation}

\subsection{Einstein Equations}
Complete system of macroscopic equations comprises of kinetic equations (\ref{kin_eq}), field equations (\ref{EqPhi}) or (\ref{EqPhiR}) and  Einstein equations:
\begin{equation}\label{Einst_Scalar}
R^{ik}-\frac{1}{2}Rg^{ik}=8\pi (T^{ik}_p+T^{ik}_s),
\end{equation}
where $T^{ik}_p$ is a defined above energy-momentum tensor of a statistical system and $T^{ik}_s$ is an energy-momentum tensor of a scalar field (\ref{Tiks}) or (\ref{TiksR}).
Covariant divergence of Einstein equations turns them into identities as a result of transport equation
(\ref{transport_Tik}) and field equation (\ref{EqPhi}) or (\ref{EqPhiR}).

\section{Cosmological Evolution Equations}
\subsection{Kinetic Equations}
Let us assume scalar potential to be function only on time $\Phi(\eta)$ for a space-flat Friedmann Universe:
\begin{equation}\label{Freedman}
ds^2=a^2(\eta)(d\eta^2-dx^2-dy^2-dz^2)\equiv a^2(\eta)ds^2_0.
\end{equation}
Let us also assume distribution function depending only on time and squared space momentum:
\begin{equation}\label{f(eta,P)}
f_a=f_a(\eta,P);\quad P_0^2=P^2_1+P^2_2+P^2_3,
\end{equation}
so that kinetic energy of particles is equal to
\begin{equation}\label{E}
E=\sqrt{P_4P^4}=\frac{P_4}{a}=\sqrt{q^2\Phi^2(\eta)+\frac{P^2_0}{a^2}}.
\end{equation}
Calculating Poisson bracket (\ref{HPsi}), let us reduce kinetic equations (\ref{kin_eq}) to the next form:
\begin{equation}\label{Puas_P4}
\frac{\partial f}{\partial \eta}=0.
\end{equation}
Thus, homogenous isotropic solution of collisionless kinetic equation in Friedmann metrics does not explicitly depend on time variable which was also true for a gas of neutral particles \cite{Yu_cosm1,Yu_cosm2}:
\begin{equation}\label{exact_kin_sol}
f(\eta,P_0)=f(P_0).
\end{equation}

\subsection{Sewing of the Collisionless Solution and Equilibrium Solution}
To find certain distribution in plasma where previously thermodynamic equilibrium was previously maintained, let us apply next method \cite{Yu_cosm1,Yu_cosm2}. Let us assume that till certain instant $\eta_0$ local thermodynamic equilibrium was maintained in plasma and it was instantly violated at the same instant. Thus, at $\eta<\eta_0$ distribution functions were locally equilibrium i.e.:
\begin{eqnarray}\label{f_LTE}
f_0(\eta,P_4)=\biggl[\exp\biggl(-\frac{\mu(\eta)}{\theta}+\nonumber\\
+\frac{\sqrt{a^2q^2\Phi^2+P^2_0}}{a\theta(\eta)}\biggr)\pm\biggr]^{-1},\quad \eta<\eta_0
\end{eqnarray}
($\mu$ are chemical potentials, $\theta$ is a temperature), and at $\eta>=\eta_0$ they were described by collisionless kinetic equations. Therefore at instant  $\eta=\eta_0$ должно иметь место равенство:
\begin{eqnarray}\label{crosslinking}
f(P_0)=\biggl[\exp\biggl(-\frac{\mu(\eta_0)}{\theta(\eta_0)}+\nonumber\\
\frac{\sqrt{q^2a^2(\eta_0)\Phi^2(\eta_0)+P^2_0)}}{\theta(\eta_0)a(\eta_0)}\biggr)\pm 1\biggr],\quad \eta=\eta_0.
\end{eqnarray}
The single option to fulfill this condition is possible if $P^2_0$ in equilibrium distribution is substituted by integral:
\begin{equation}\label{P0->P0}
\mathcal{P}^2_0(\eta,P_0)= P_0^2-\int\limits_{\eta_0}^\eta  a^2\frac{d\Phi^2}{d\eta}d\eta.
\end{equation}
Thus, the exact solution of collisionless kinetic equations(\ref{kin_eq}), which becomes equilibirum one at $\eta=\eta_0$, is:
\newcommand{\fo}[1]{\frac{#1}{\exp\biggl(-\frac{\mu(\eta_0)}{\theta_0}
+\frac{\sqrt{m^2_0+p^2}}{\theta_0}\biggr)\pm 1}}
\begin{eqnarray}\label{exact_f_LTE}
f(\eta,P_0)=\biggl[\exp\biggl(-\frac{\mu_0}{\theta_0}+\frac{\sqrt{m^2_0+p^2}}{\theta_0}\biggr)\pm 1\biggr]^{-1},
\end{eqnarray}
where the following denotations are introduced:
\begin{eqnarray}\label{tildem}
m_0=a_0q\Phi_0;\; a_0=a(\eta_0);\; \mu_0=\mu(\eta_0);\\
\theta_0=\theta(\eta_0);\; \Phi_0=\Phi(\eta_0);\; P_0=a_0 p.
\end{eqnarray}
Thus, $m_0$ is an effective particle mass at instant of equilibrium violation. Further for the sake of brevity we will call the solution (\ref{exact_f_LTE}) {\it quasi-equilbrium solution}.

\subsection{Moments of Quasi-equilibrium Distribution Function}
It is easily seen that quasi-equilibrium distribution function's moments (\ref{cons_n}) and (\ref{Tpl}) which are interesting for us, coincide in the algebraic structure with corresponding moments of locally-equilibrium distribution:
\begin{eqnarray}\label{ni}
n^i=u^i n; \quad u^i=\frac{1}{a}\delta^i_4;\\
\label{Tik}
T^{ik}_{pl}=(\mathcal{E}_{pl}+P_{pl})u^iu^k-P_{pl}g^{ik},
\end{eqnarray}
where $u^i$ is a unit timelike vector of a synchronous observer. However, macroscopic scalars $n,P,\mathcal{E},\sigma$ relative to quasi-equilibrium distribution (\ref{exact_f_LTE}) do not coincide with corresponding scalars relative to equilibrium distribution function.
Calculating, we find\footnote{$\gamma=\mu/\theta$ which is a reduced chemical potential}:
\begin{equation}\label{n}
n=\frac{\rho a^3_0}{2\pi^2a^3(\eta)}\int\limits_0^\infty \fo{p^2dp},
\end{equation}
so that:
\begin{equation}\label{na3_const}
na^3=\Const,
\end{equation}
which ensures conservation law of particle number. Thus, the expression for particle number density coincides with standard equilibrium one with constant mass, temperature and chemical potential within the accuracy of multiplication by scale factor $(a_0/a)^3$.

Expressions for remaining three scalars do not coincide with equilibrium ones:
\begin{equation}
\label{E}
\mathcal{E}_{pl} =\biggl( \frac{a_0}{a}\biggr)^4 \frac{\rho}{2\pi^2}
 \int\limits_{0}^{\infty} \fo{p^2\sqrt{\bar{m}^2 +p^2} d p}\,;
 \end{equation}
 \begin{eqnarray}
\label{P}
P_{pl} = \biggl(\frac{a_0}{a}\biggr)^4\frac{\rho}{6\pi^2} \times \nonumber\\
\hskip -12pt \int\limits_{0}^{\infty} \frac{p^4}{\sqrt{\bar{m}^2+p^2}}\fo{dp}\,; \\
\label{sigma_0}
\sigma =\Phi \biggl(\frac{a_0}{a}\biggr)^2\frac{\rho q^2}{2\pi^2} \times \nonumber\\
\hskip -12pt \int\limits_{0}^{\infty} \frac{p^2
}{\sqrt{\bar{m}^2+p^2}}\fo{dp}\,,
\end{eqnarray}
where:
\begin{equation}\label{m*m0}
\bar{m}=m_*\frac{a}{a_0}\equiv \frac{a\Phi}{a_0\Phi_0}m_0=\bar{m}(\bar{\Phi}).
\end{equation}
The behavior of these scalars is substantially defined by the behavior of invariant:
\begin{equation}\label{af}
\bar{\Phi}\equiv a(\eta)\Phi(\eta).
\end{equation}
Let us note that at $\bar{\Phi}=\Const$ integrals in expressions for all three scalars are constant values.

\subsection{Field Equations}
In case of conformally non-invariant field (\ref{Ls}) field equation (\ref{EqPhi}) in Friedmann metrics takes form:
\begin{equation} \label{EqS_Freed}
\frac{1}{a^4}\frac{d}{d\eta}a^2\frac{d}{d\eta}\Phi+\epsilon_2 m^2_s\Phi=-4\pi\epsilon_1\sigma.
\end{equation}
In case of conformally invariant scalar field (\ref{LsR}) this equation needs to be substituted by the following one:
\begin{equation} \label{EqSR_Freed}
\frac{1}{a^3}\frac{d^2}{d\eta^2}a\Phi+\epsilon_2 m^2_s\Phi=-4\pi\epsilon_1\sigma.
\end{equation}

Let us consider conformally invariant case of {\it massless} scalar field. Let us rewrite scalar charge density
(\ref{sigma_0}) in the following form:
\begin{equation}\label{sigma_trans}
\sigma=\frac{\bar{\sigma}(\bar{\Phi})}{a^3},
\end{equation}
where
\begin{eqnarray}\label{barsigma}
\bar{\sigma}(\bar{\Phi}) =\bar{\Phi} \frac{\rho q^2a_0^2}{2\pi^2} \times \nonumber\\
\hskip -12pt \int\limits_{0}^{\infty} \frac{p^2
}{\sqrt{\bar{m}^2+p^2}}\fo{dp}\,,
\end{eqnarray}
Thus, conformally invariant equation (\ref{EqSR_Freed}) for a massless scalar field takes form:
\begin{equation} \label{EqSR_Freed}
\frac{d^2}{d\eta^2}\bar{\Phi}=-4\pi\epsilon_1\bar{\sigma}(\bar{\Phi}).
\end{equation}
i.e. equation for scalar conformally invariant massless field with a source becomes autonomous ordinary second-order integro-differential equation.
 We can find the solutions of this equation in different extreme cases.

Let us first consider a {\it ultrarelativistic case}:
\begin{equation}\label{ultra-ultra}
\langle p\rangle\gg \mathrm{max}(m_0,\bar{m}).
\end{equation}
In such a case equation (\ref{EqSR_Freed}) becomes a linear second-order differential equation with constant coefficients:
\begin{equation}\label{osc_f}
\frac{d^2}{d\eta^2}\bar{\Phi}+\epsilon_1 \omega^2_0\bar{\Phi}=0,
\end{equation}
where
\begin{equation}\label{omega0}
\omega^2_0=\frac{\rho q^2a_0^2}{2\pi^2}
\int\limits_{0}^{\infty} \frac{pdp
}{\exp(\frac{-mu_0+p}{\theta_0})\pm 1}\,.
\end{equation}
In a case of like-charged particles repulsion ($e_1=+1$) this equation has its general solution:
\begin{equation}\label{fe1+1}
\Phi= C_1\frac{\cos(\omega_0\eta)}{a(\eta)}+C_2\frac{\sin(\omega_0\eta)}{a(\eta)},
\end{equation}
and in a case of attraction ($e_1=-1$) it is:
\begin{equation}\label{fe1-1}
\Phi= C_1\frac{\cosh(\omega_0\eta)}{a(\eta)}+C_2\frac{\sinh(\omega_0\eta)}{a(\eta)}.
\end{equation}
Let us notice that exactly the same solution was found in \cite{conform} for ultrarelativistic equilibrium plasma. This coincidence is easily  explained
since quasi-equilibrium distribution (\ref{exact_f_LTE}) coincides with equilibrium one in the ultrarelativistic limit.

Let us consider now a {\it non-relativistic case}\footnote{Let us note that here the smallness of average momentum is required relative to $\bar{m}$ but not relative to $m_0$.}:
\begin{equation}\label{nonultra-nonultra}
\langle p\rangle\ll \bar{m}.
\end{equation}
Then instead of homogenous equation (\ref{osc_f}) we get inhomogeneous equation:
\begin{equation}\label{nonosc_f}
\frac{d^2}{d\eta^2}\bar{\Phi}=-4\pi\epsilon_1 \sigma_0,
\end{equation}
where
\begin{equation}\label{sigma0}
\sigma_0=\frac{\rho q^2a_0\Phi_0}{2\pi^2 m_0}
\int\limits_{0}^{\infty} \frac{pdp
}{\exp\bigl(\frac{-mu_0+\sqrt{m^2_0+p^2}}{\theta_0}\bigr)\pm 1}\,,
\end{equation}
The solutions of this equation are:
\begin{equation}\label{phi_non}
\Phi=\frac{C_1}{a}+C_2\frac{\eta}{a} -2\pi\epsilon_1 \sigma_0\frac{\eta^2}{a}.
\end{equation}
\subsection{Einstein Equation}
It can easily be seen that energy-momentum tensor of conformally invariant massless scalar field $\Phi(\eta)$ has a form of energy-momentum tensor of the ideal flux ($e_1=+1$)
\begin{equation} \label{TiksRa_} T_{(s)}^{ik} =(\mathcal{E}_{s} +P_{s} )v^{i} v^{k} -P_{s} g ^{ik} , \end{equation}
where
\begin{equation} \label{EsRa}
\mathcal{E}_{s} =\frac{1}{8\pi a^4}\bar{\Phi}'^2; \end{equation}
\begin{equation} \label{PsRa} P_{s} =\frac{1}{24\pi a^4} \left[ \bar{\Phi}'^2
+8\pi\epsilon_1\bar{\Phi}\bar{\sigma}(\bar{\Phi})\right]. \end{equation}
so that:
\begin{equation} \label{TsRa}
T_{s} =\mathcal{E}_{s} -3P_{s} =\frac{1}{a^4}\bar{\Phi}\bar{\sigma}(\bar{\Phi}).
\end{equation}
The single non-trivial Einstein equation in Friedmann metrics takes the next form:
\begin{equation}\label{Einst_Freed}
\frac{a'^2}{a^4}=\frac{8\pi}{3}(\mathcal{E}_s+\mathcal{E}_{pl}).
\end{equation}
Substituting the expressions for energy density from (\ref{E}) and (\ref{EsRa}) into this equation, we reduce it to:
\begin{equation}\label{Eq_Einst_conf}
a'^2=\frac{1}{3}\bar{\Phi}'^2+\frac{8\pi}{3}\bar{\mathcal{E}}_{pl}(\bar{\Phi}),
\end{equation}
\begin{equation}\label{barE}
\bar{\mathcal{E}}_{pl}(\bar{\Phi})=\frac{a_0^4\rho}{2\pi^2}
 \int\limits_{0}^{\infty} \fo{p^2\sqrt{\bar{m}^2 +p^2} d p}.
\end{equation}
Thus, cosmological model in case of massless conformally invariant scalar field with a collisionless source is reduced to two ordinary integro-differential equations of the second order (\ref{EqSR_Freed}) and first order (\ref{Eq_Einst_conf}) relative to functions $a(\eta)$ and $\bar{\Phi}(\eta)$.

In particular, for the ultrarelativistic case we have $\bar{\mathcal{E}}_{pl}(\bar{\Phi})=\bar{\mathcal{E}}_0=\Const$. Choosing a convergent solution of scalar field equation (\ref{fe1+1})
\begin{equation}\label{barPhiultra}
\bar{\Phi}=\phi_0\sin(\omega_0\eta)
\end{equation}
and substituting it into Einstein equation (\ref{Eq_Einst_conf}), we reduce it to the next form
\begin{equation}\label{Einst_Freed}
a'^2=\frac{8\pi}{3}\phi^2_0\omega^2_0\cos^2\omega_0\eta+\frac{8\pi}{3}\bar{\mathcal{E}}_0.
\end{equation}
The solution of this ordinary differential equation has the following form:
\begin{eqnarray}\label{solve_a}
a(\eta)=\sqrt{\frac{8\pi\bar{\mathcal{E}}_0}{3|\bar{\Phi}_0|}}\mathrm{Re}(\mathrm{E}(\omega_0\eta,k)),\\
\label{k}
k^2=\frac{\alpha^2}{\alpha^2+\beta^2};\quad \alpha^2=\phi^2_0\omega^2_0;\; \beta^2=\bar{\mathcal{E}}_0,
\end{eqnarray}
$\mathrm{E}(x,k)$ is an elliptic integral of the second kind:
\begin{equation}
\mathrm{E}(\omega_0\eta,k)=\int\limits_0^x\sqrt{1-\sin^2x}dx.
\end{equation}

\section*{Acknowledgements}
This work was founded by the subsidy allocated to Kazan Federal University for the state assignment in the sphere of scientific activities.
Autor expresses his gratitude to participants of MW seminar for relativistic kinetics and cosmology of Kazan Federal University for helpful discussion of the work.
%%%%%%%%%%%%%%%%%%%%%%%%%%%%%%%%%%%%%%%%%%%%%%%%%%%%%%%%%%%%%%%%%

\end{document}